\documentclass[12pt]{iopart}

\usepackage{graphicx}
\usepackage{csquotes}
\MakeOuterQuote{"}
\usepackage{bm}
\usepackage{hyperref}
\hypersetup{
    unicode=false,  
    pdftoolbar=true,
    pdfmenubar=true,
    pdffitwindow=false,
    pdfstartview={FitH},
    pdftitle={On the magnetic moment of a runaway electron},
    pdfauthor={Chang Liu},
    pdfsubject={Guiding-center theory},
    pdfkeywords={Guiding center} {Runaway electron} {Magnetic moment},
    pdfnewwindow=true,
    colorlinks=true,
    linkcolor=blue,
    citecolor=blue,
    filecolor=blue,
    urlcolor=blue
}
\usepackage[numbers,square,sort&compress]{natbib}
\usepackage{color}

\begin{document}
	
\title[Magnetic moment and collisionless pitch-angle scattering]{The role of magnetic moment in the collisionless pitch-angle scattering of runaway electrons}

\author{Chang Liu}
\address{Princeton Plasma Physics Laboratory, Princeton, New Jersey 08540, USA}
\ead{cliu@pppl.gov}
\author{Hong Qin}
\address{School of Physics, Princeton Plasma Physics Laboratory, Princeton, New Jersey 08540, USA}
\address{University of Science and Technology of China, Hefei, Anhui, 230000, China}
\author{Eero Hirvijoki}
\address{Princeton Plasma Physics Laboratory, Princeton, New Jersey 08540, USA}
\ead{ehirvijo@pppl.gov; eero.hirvijoki@gmail.com}
\author{Yulei Wang}
\address{School of Physics, University of Science and Technology of China, Hefei, Anhui, 230000, China}
\author{Jian Liu}
\address{School of Physics, University of Science and Technology of China, Hefei, Anhui, 230000, China}

\begin{abstract}
Recently, the validity of the guiding-center approach to model relativistic runaway electrons in tokamaks has been challenged by full-orbit simulations that demonstrate the breakdown of the standard magnetic moment conservation. In this paper, we derive a new expression for the magnetic moment of relativistic runaway electrons, which is conserved significantly better than the standard one. The new result includes one of the second-order corrections in the standard guiding-center theory which, in case of runaway electrons with $p_{\parallel}\gg p_{\perp}$, can peculiarly be of the same order as the lowest-order term. The better conservation of the new magnetic moment also explains the collisionless pitch-angle-scattering effect observed in full-orbit simulations since it allows momentum transfer between the perpendicular and parallel directions when the runaway electron is accelerated by an electric field. While the derivation of the second-order correction to the magnetic moment in general case would require the full extent of the relativistic second-order guiding-center theory, we exploit the Lie-perturbation method at the limit $p_{\parallel}\gg p_{\perp}$ which simplifies the computations significantly. Consequently, we present the corresponding  guiding-center equations applicable to the highly relativistic runaway electrons.
\end{abstract}

\maketitle

\section{Introduction}

When the energy of an electron in a plasma surpasses a certain threshold, a strong enough parallel electric field may accelerate it to extremely high energies, due to the fact that the collisional drag force decreases with the increasing electron energy~\cite{dreicer_electron_1959,dreicer_electron_1960,connor_relativistic_1975}. Such electrons are called runaway electrons (REs)~\cite{helander_runaway_2002}, and have been found in tokamak experiments during the startup~\cite{jaspers_experimental_1993} and flat-top~\cite{zhou_investigation_2013,paz-soldan_spatiotemporal_2017} phases, and especially during disruptions~\cite{gill_generation_1993,lehnen_disruption_2011}. It is predicted that in a large tokamak device like ITER, a large population of runaway electrons could be generated in a typical disruption event, with energies in tens of MeVs~\cite{Jayakumar-1993,rosenbluth_putvinski_1997,boozer_theory_2015,martin-solis_formation_2017}. Such runaway electrons have the potential to cause significant damage to the device, raising one of the key questions for planning disruption mitigation in ITER. It is thus critical to have physical understanding and reliable simulation tools to model the highly relativistic runaway electrons in a tokamak geometry.

In strongly magnetized plasma, the motion of a runaway electron consists of rapid gyration around the magnetic field line, fast parallel motion along the field line, and slow drift across the field lines.
For the sake of numerical efficiency, it would be desirable to apply the so-called guiding-center~\cite{northrop_adabatic_1963,littlejohn_variational_1983,cary_hamiltonian_2009} approximation which decouples the rapid gyro-motion from the parallel and drift motions. %~\cite{guan_phase-space_2010,liu_what_2014,papp_runaway_2011}.
Recent studies, however, hint that the guiding-center approximation might not necessarily be valid for tracing the highly relativistic REs ~\cite{liu_collisionless_2016,carbajal_space_2017}. By tracking the full orbit of electrons in 6-D phase-space, it has been claimed that for highly relativistic REs, the magnetic moment $\mu$ would not remain an adiabatic invariant, even in the absence of dissipative effects. In the literature this phenomena is referred to as "collisionless pitch-angle scattering".
It has been argued that for runaway electrons with $v_{\parallel}\sim c$ and $\gamma_{r}\gg 1$ ($v_{\parallel}$ is the parallel velocity, $c$ is the speed of light, and $\gamma_{r}$ is the Lorentz factor), due to the curved magnetic field in tokamak geometry, the variation of the magnetic field $\mathbf{B}$ along the electron trajectory within one gyro-period is not small, hence breaking the assumption behind the guiding-center approximation. However, according to the simulation results presented in \cite{liu_collisionless_2016}, the breakdown of $\mu$ conservation appears to happen even if $\gamma_{r}$ is not very large and the conditions for guiding-center approximation could still be regarded as valid.

In this paper, we address the mystery described above and demonstrate that relativistic runaway electrons still display a magnetic moment that can be regarded as a good adiabatic invariant over sufficiently long periods of time. The corrections that we derive depend not only on the momentum perpendicular to the magnetic field ($\mathbf{p}_{\perp}$), but also on the parallel momentum ($p_{\parallel}$) and the magnetic field-line curvature vector ($\bm{\kappa}$). To understand the origin of these corrections, we revisit the Hamiltonian guiding-center theory and the derivation of the relativistic guiding-center phase-space Lagrangian using the non-canonical Lie-perturbation method~\cite{cary_littlejohn_1983,littlejohn_variational_1983,brizard_foundations_2007}. In our derivation, in addition to the standard guiding-center ordering ($\rho_{\parallel}\ll L$, $\rho_{\perp}\ll L$, where $\rho_{\parallel}=p_{\parallel}/qB$, $\rho_{\perp}=p_{\perp}/qB$, $q$ is the charge, and $L$ is the scale length of magnetic field inhomogeneity), we introduce the condition $p_{\perp}\ll p_{\parallel}$. %to reflect the fact that REs typically display a strong anisotropy in the momentum~\cite{landreman_numerical_2014,aleynikov_theory_2015,liu_role_2018}. 
This assumption greatly simplifies the derivation of the guiding-center theory, and lets us carry it trough to second order with not much trouble. Were the ordering $p_{\perp}\ll p_{\parallel}$ not introduced, one would be forced to carry out the second-order guiding-center theory to full extent which is an intimidating task only a handful of authors have ever engaged upon~\cite{brizard_equivalent_2012,burby_automation_2013,tronko_lagrangian_2015,brizard_equivalent_2016}, and none in the relativistic case. With the help of the additional ordering assumption, we find a new relativistic guiding-center phase-space Lagrangian and the corresponding equations of motion that decouple the gyro-motion from the parallel and drift motions. Most importantly, our derivation underlines that the root cause for the observed collisionless pitch-angle scattering most likely is the existence of an adiabatically invariant $\mu$, in contrary to what has been previously proposed: our expression provides a channel for momentum transfer between parallel and perpendicular directions that matches the full-orbit simulations quite well. Hence the presented work paves the road for better understanding of the simulations of relativistic runaway electrons in tokamaks.

The rest of this paper proceeds in a following manner. In Sec.~\ref{sec:numerical-indication}, we present the new expression for the magnetic moment of relativistic runaway electrons, use full-orbit simulations to demonstrate that it is conserved significantly better than the standard expression, and explain the collisionless pitch-angle-scattering effect as a consequence of it. In Sec.~\ref{sec:guiding-center-transformation}, we follow the Lie-perturbation method to derive the guiding-center transformation for relativistic REs assuming $p_{\perp}\ll p_{\parallel}$, providing the corresponding phase-space Lagrangian. In Sec.~\ref{sec:equation-of-motion}, we derive the associated guiding-center Poisson bracket and the equations of motion and, in Sec.~\ref{sec:higher-order-terms}, we discuss how the magnetic moment introduced in Sec.~\ref{sec:numerical-indication} is consistent with the second-order nonrelativistic guiding-center theory at the limit $p_{\perp}\ll p_{\parallel}$. Finally, Sec.~\ref{sec:conclusion} concludes our work.

\section{Asymptotic invariance of magnetic moment and collisionless pitch-angle scattering}
\label{sec:numerical-indication}

As shown in \cite{liu_collisionless_2016,carbajal_space_2017}, for relativistic runaway electrons in tokamaks, the standard expression for the magnetic moment, $\mu_0=p_{\perp}^{2}/2mB$, does not remain a good adiabatic invariant. In fact, it was shown that, as the electron is accelerated, $\mu_0$ can grow to 100 times its original value, displaying strong oscillations with a timescale corresponding to the gyroperiod. This was considered as an indication of the breakdown of the magnetic moment and the standard guiding-center theory. The story, however, is slightly more complicated. %The standard first order guiding-center theory, however, provides also the first order correction to the magnetic moment, this expression was not explored in the numerical simulations. Nevertheless, since also the first order correction is proportional to the perpendicular momentum, we could expect the anisotropy issue to remain, as will be demonstrated soon. Only after including the second order correction term that is independent of perpendicular momentum, do we find a quantity that is preserved for sufficiently long.

We start by introducing a new expression for the magnetic moment of relativistic runaway electrons (this will be derived later using Lie-perturbation methods). The expression consists of three terms
\begin{eqnarray}\label{eq:mu}
\mu &=\frac{|\mathbf{p}_{\perp}+p_{\parallel}^{2}\bm{\kappa}\times\mathbf{b}/(qB)|^{2}}{2mB},\nonumber\\
 &=\mu_0+\mu_1+\mu_2,
\end{eqnarray}
\begin{equation}\label{eq:mu-separate}
  \mu_{0}=\frac{|\mathbf{p}_{\perp}|^2}{2mB},\qquad \mu_{1}=\frac{p_{\parallel}^{2}\mathbf{p}_{\perp}\cdot\bm{\kappa}\times \mathbf{b}}{qmB^{2}},\qquad \mu_{2}=\frac{p_{\parallel}^4}{q^2B^2}\frac{|\bm{\kappa}\times\mathbf{b}|^2}{2mB},
\end{equation}
where $\mathbf{b}=\mathbf{B}/|B|$ is the magnetic field unit vector, $\mathbf{p}_{\perp}=\mathbf{b}\times(\mathbf{p}\times \mathbf{b})$ is the kinetic momentum perpendicular to $\mathbf{b}$, $p_{\parallel}=\mathbf{p}\cdot\mathbf{b}$ is the momentum parallel to $\mathbf{b}$, and $\bm{\kappa}=\mathbf{b}\cdot\nabla\mathbf{b}$ is the magnetic field-line curvature vector. The term $\mu_1$ is familiar from the relativistic first-order guiding-center theory and we expect that $\mu_2$ would be found the same way lurking at second order. It is, however, a devious task to carry out the second order theory to full extent. The only second order expression for $\mu$ we are aware of was derived using the guiding-center automation algorithm~\cite{burby_automation_2013} in the nonrelativistic case. Assuming that the expression in~\cite{burby_automation_2013} would generalize to the relativistic case, it would coincide with our $\mu_2$ assuming the additional ordering $p_{\perp}\ll p_{\parallel}$. Note that while our expression was derived for a relativistic case, it is strictly valid only if $p_{\perp}\ll p_{\parallel}$ is assumed.
%For thermal electrons or ions with $p_{\parallel}\sim \mathbf{p}_{\perp}$, the $\mu_2$ is one order of magnitude smaller than $\mu_2$, given that $p_{\parallel}\bm{\kappa}\times \mathbf{b}/(qB)\sim \rho_{\parallel}/L\ll 1$ but one should use the full expression first order correction from the standard guiding-center theory instead. For REs with $p_{\parallel}\gg p_{\perp}$, the second term may become comparable or larger than $\mathbf{p}_{\perp}$ the lowest-order expression first one.

To illustrate the behavior of the new $\mu$, we carry out full-orbit simulations with an advanced phase-space volume-conserving algorithm~\cite{liu_collisionless_2016}. The simulations are conducted with parameters similar to the EAST tokamak: major radius $R_0=1.7\,\textrm{m}$, safety factor $q=2$, central magnetic field $B_{0}=3\, \textrm{T}$. The initial parallel and perpendicular momentum of the test runaway electron are set to $p_{\parallel 0}=5mc$ and $p_{\perp 0}=mc$ respectively, and the toroidal electric field is $E=0.2\, \textrm{V/m}$. The initial position of the test electron is $R=1.8\, \textrm{m}$ and $Z=0.0\, \textrm{m}$, which is close to the magnetic axis. 

\begin{figure}[h]
	\begin{center}
		\includegraphics[width=0.9\linewidth]{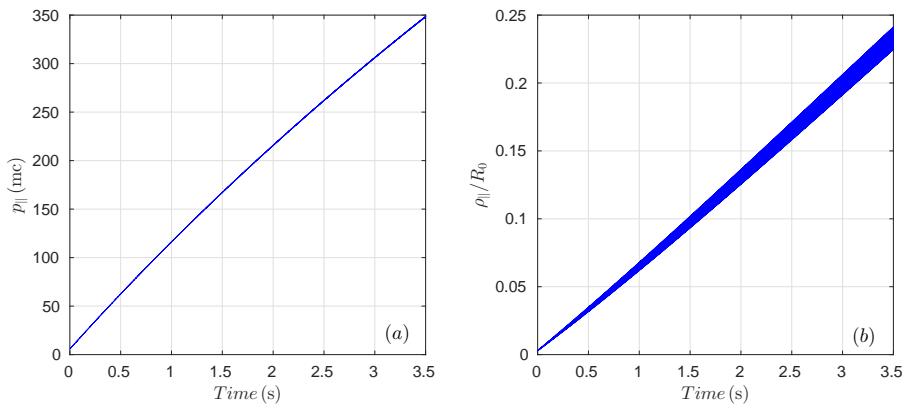}
	\end{center}
	\caption{\label{fig:time-evolution} The time evolution of $p_{\parallel}$ (a) and $\rho_{\parallel}/R_{0}$ (b) for the test electron.}
\end{figure}

In Fig.~\ref{fig:time-evolution}, the panel (a) depicts the time evolution of the test runaway electron's parallel momentum $p_{\parallel}$, which keeps growing in time due to the electric field acceleration, and panel (b) shows the corresponding evolution of $\rho_{\parallel}/R_{0}$, which characterizes the validity of the guiding-center approximation. As is clear, the quantity $\rho_{\parallel}/R_{0}$ grows large, indicating that the standard guiding-center theory likely breaks down, and that higher-order contributions are necessary to recover valid asymptotic guiding-center motion and an adiabatic invariant corresponding to $\mu$. %For $t<1.5$s, we find that $\rho_{\parallel}/R_{0}$ is smaller than 0.1, and in the later time this parameter gets larger as $p_{\parallel}$ increases.

\begin{figure}[h]
	\begin{center}
		\includegraphics[width=0.9\linewidth]{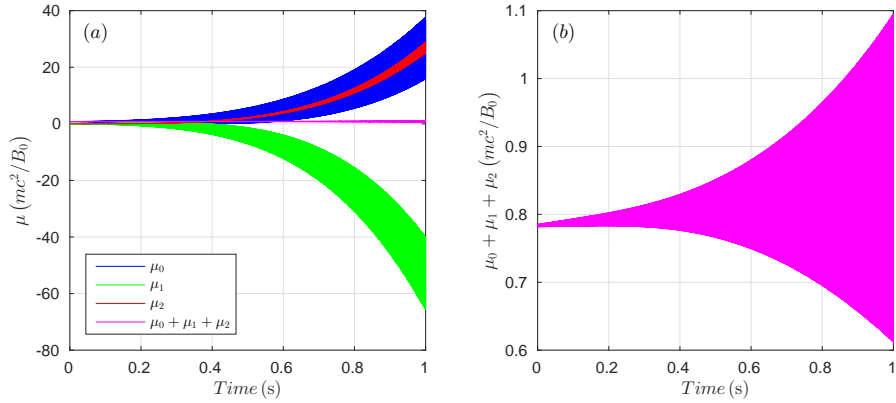}
	\end{center}
	\caption{\label{fig:mu-conservation1s}(a) The time evolution of $\mu_{0}$, $\mu_1$, $\mu_2$, and the new magnetic moment $\mu$ which is the sum of them, from 0 to 1s. (b) A zoomed-in of the time evolution of $\mu$.}
\end{figure}

\begin{figure}[h]
	\begin{center}
		\includegraphics[width=0.9\linewidth]{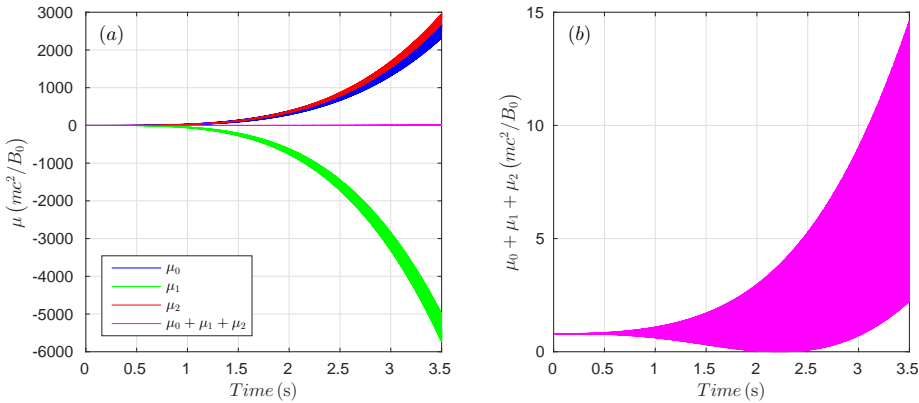}
	\end{center}
	\caption{\label{fig:mu-conservation3s}(a) The time evolution of $\mu_{0}$, $\mu_1$, $\mu_2$, and $\mu$, from 0 to 3.5s. (b) A zoomed-in of the time evolution of $\mu$.}
\end{figure}

The evolution of $\mu$, as defined in Eqs.~(\ref{eq:mu}-\ref{eq:mu-separate}), is illustrated in Fig.~\ref{fig:mu-conservation1s} and Fig.~\ref{fig:mu-conservation3s} until $1.0\,\textrm{s}$ and $3.5\,\textrm{s}$ respectively, with contributions from $\mu_0$, $\mu_1$, and $\mu_2$ separated. The standard magnetic moment $\mu_0$ clearly is not conserved at all and experiences strong oscillations and drift, in agreement with the previous studies presented at~~\cite{liu_collisionless_2016}. The expression $\mu=\mu_0+\mu_1+\mu_2$, however, experiences orders of magnitudes smaller deviations and drift from its original value. %In case of runaway electrons with energy less than 160 $mc^{2}$ ($\sim$ 82 MeV), the expression $\mu=\mu_0+\mu_1+\mu_2$ could still be considered as an invariant. 
Nevertheless, if the simulation is carried until several seconds, also this expression starts to oscillate and drift significantly as  $\rho_{\parallel}/R_{0}$ becomes larger and breaks the second-order guiding-center theory. As the sequence $\mu_0,\mu_0+\mu_1,\mu_0+\mu_1+\mu_2$ clearly displays convergence, the simulations indicate that an asymptotic invariant could well exist but one would need to use even higher-order theory to find an expression valid for times scales of seconds.

To explain the observed pitch-angle-scattering effect, we first note that the quasi-conservation of the higher-order magnetic moment provides a channel for converting parallel momentum to perpendicular momentum. As indicated in Eq.~(\ref{eq:mu}), $\mu$ depends on both $\mathbf{p}_{\perp}$ and $p_{\parallel}$. Given that for REs, $p_{\parallel}$ will keep growing because of the electric field acceleration, then, to make $\mu$ an invariant, $p_{\perp}$ will also grow. Furthermore, we can solve Eq.~(\ref{eq:mu}) for $p_{\perp}$ in terms of $p_{\parallel}$ and $\mu$ according to
\begin{equation}
\label{eq:pperp-ppar}
  p_{\perp}=\sqrt{\left(p_{\parallel}^{2}\frac{\bm{\kappa}\times \mathbf{b}}{qB}\right)^2+2\mu m B+p_{\parallel}^{2}\frac{|\bm{\kappa}\times \mathbf{b}|}{qB}\sqrt{2\mu m B}\cos\theta},
\end{equation}
where $\theta$ is the angle between $\bm{\kappa}\times\mathbf{b}$ and $\mathbf{p}_{\perp}+p_{\parallel}^{2}\bm{\kappa}\times \mathbf{b}/(qB)$. If we ignore the cross product term assuming it is an oscillatory term, we can get an approximate value $\mathbf{p}'$,
\begin{equation}
\label{eq:pperp-prime}
p'_{\perp}=\sqrt{\left(p_{\parallel}^{2}\frac{\bm{\kappa}\times \mathbf{b}}{qB}\right)^2+2\mu m B}.
\end{equation}

\begin{figure}[h]
	\begin{center}
		\includegraphics[width=0.5\linewidth]{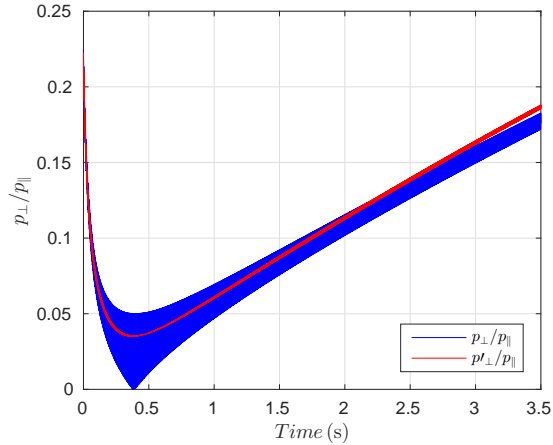}
	\end{center}
	\caption{\label{fig:scattering}The change of $p_{\perp}/p_{\parallel}$ and $p'_{\perp}/p_{\parallel}$ with time for the test electron.}
\end{figure}

The value of $p_{\perp}/p_{\parallel}$ and $p'_{\perp}/p_{\parallel}$ as functions of time from the full-orbit simulation is plotted in Fig.~\ref{fig:scattering}, which characterize the pitch-angle of the electrons. Despite the oscillatory part, the values of these two quantities are very close even after the guiding-center ordering breaks down. According to the results shown in Fig.~\ref{fig:scattering}, the increase of $p_{\perp}$ and pitch-angle from the full-orbit simulation can be explained through the conservation of $\mu$ in Eq.~(\ref{eq:mu}) to a large extent. This gives an explanation of the collisionless pitch-angle scattering observed in~\cite{liu_collisionless_2016,carbajal_space_2017}. This scattering effect can be important for the dynamics of runaway electrons, including their radiation effects, the energy distribution and the coupling to plasma MHD behaviors.

\section{Guiding-center transformation for relativistic runaway electrons}
\label{sec:guiding-center-transformation}

In this section we introduce a guiding-center transformation for relativistic runaway electrons. The transformation is based on the non-canonical Hamiltonian mechanics approach using Lie perturbation method~\cite{cary_littlejohn_1983,littlejohn_variational_1983,brizard_foundations_2007}, which is based on the guiding-center ordering. By doing an infinitesimal transformation of the coordinates through a generating function $\mathbf{G}$, we manage to remove the rapid gyro motion from both the Lagrangian and the Hamiltonian. We will show that in order to achieve such transformation for runaway electrons with $p_{\perp}\ll p_{\parallel}$, a new set of guiding-center coordinates needs to be introduced, including the new magnetic moment $\mu$ and the corresponding gyro phase angle $\theta$, which is different from those in the standard guiding-center theory.

We begin with the Lagrangian in the local particle coordinates ($\mathbf{x}, \mathbf{p}$). The Poincar\'{e}-Cartan-Einstein%~\cite{qin_geometric_2007}
 phase-space-time one-form for a charged particle in the magnetic field can be written as
\begin{equation}
\gamma = q \mathbf{A}\cdot  d\mathbf{x} + \mathbf{p}\cdot  d\mathbf{x} - w dt,
\end{equation}
where $\mathbf{A}$ is the vector potential of the magnetic field $\mathbf{B}$. $\mathbf{p}$ is the kinetic momentum of the particle. $w$ is the particle energy. We can also write the relativistic particle Hamiltonian $h$ as~\cite{cary_hamiltonian_2009},
\begin{equation}
  h=mc^{2}\sqrt{1+p^{2}/m^{2}c^{2}}-w.
\end{equation}

Since the guiding-center transformation is an infinitesimal transformation, we need to introduce the a small parameter $\epsilon$. According to guiding-center ordering, $\rho_\parallel/L\ll 1$, $\rho_\perp/L\ll 1$. In addition, for many of the runaway electrons, the condition $p_{\parallel}\gg p_{\perp}$ is true. Therefore, we set
\begin{equation}
\label{eq:ordering}
  \frac{\rho_\parallel}{L}=\epsilon,\qquad \frac{p_{\perp}}{p_{\parallel}}=\epsilon.
\end{equation}
A result from the above ordering is
\begin{equation}
  \frac{\rho_\perp}{L}=\epsilon^{2}.
\end{equation}
The reason we use the same $\epsilon$ for both ratios are as follows. As shown in Eq.~(\ref{eq:pperp-ppar}), for RE with momentum dominated by $p_{\parallel}$, the second and the third terms on the right-hand-side are ignorable compared to the first one. In this case the value of $p_{\perp}/p_{\parallel}$ is the same of $\rho_{\parallel}/L$. This can also be illustrated by comparing Fig.~\ref{fig:time-evolution} (b) and Fig.~\ref{fig:scattering} for the later time part.

The phase-space-time one-form can then be expanded as
\begin{equation}
\gamma = \gamma_{0} + \epsilon\gamma_{1} + \epsilon^{2}\gamma_{2} + \dots,
\end{equation}
where
\begin{equation}
  \gamma_{0} =q \mathbf{A}\cdot d\mathbf{x}+ p_{\parallel} \mathbf{b}\cdot  d\mathbf{x}-wdt,
\end{equation}
\begin{equation}
\gamma_{1} =\mathbf{p}_{\perp} \cdot  d\mathbf{x},
\end{equation}
\begin{equation}
\gamma_{2} =0.
\end{equation}
%\eero{Dimensional analysis shows that the initial ordering should be $q\bm{A}\cdot d\bm{x}+\epsilon p_{\parallel}d\bm{x}$}

Similarly, the Hamiltonian can be expanded as
\begin{equation}
h= h_{0} + \epsilon h_{1} + \epsilon^{2} h_{2} + \dots,
\end{equation}
where
\begin{equation}
h_{0} =mc^{2}\sqrt{1+p_{\parallel}^{2}/m^{2}c^{2}}-w,
\end{equation}
\begin{equation}
h_{1} =0,
\end{equation}
\begin{equation}
h_{2} =\frac{p_{\perp}^{2}}{2m\sqrt{1+p_{\parallel}^{2}/m^{2}c^{2}}}.
\end{equation}

We now use the Lie perturbation method to find the guiding-center transformation and the new phase-space-time one-form. The Lie-transform push-forward operator is defined as,
\begin{equation}
T^{-1}=\exp(-\sum_{n}\epsilon^{n}\mathcal{L}_{\mathbf{G}_{n}})=1-\epsilon \mathcal{L}_{\mathbf{G}_{1}}-\epsilon^{2}L_{\mathbf{G}_{2}}+\frac{1}{2}\epsilon^{2}L^{2}_{\mathbf{G}_{1}}+\dots
\end{equation}
where $\mathcal{L}_{\mathbf{G}_{n}}$ is the Lie-derivative generated by a vector field $\mathbf{G}_{n}$. The Lie derivative of a one-form can be calculated as
\begin{equation}
  \mathcal{L}_{\mathbf{G}}\gamma=i_{\mathbf{G}}\cdot d\gamma+d(i_\mathbf{G}\cdot \gamma),
\end{equation}
where $i$ is the contraction operator and $d$ is the exterior derivative.

The guiding-center transformation of the phase-space-time one-form and the Hamiltonian is
\begin{equation}
  \Gamma=T^{-1}\gamma=\Gamma_{0}+\epsilon\Gamma_{1}+\epsilon^{2}\Gamma_{2}
\end{equation}
\begin{equation}
H=T^{-1} h=H_{0}+\epsilon H_{1}+\epsilon^{2} H_{2}
\end{equation}

The expressions for the first order terms are simply,
\begin{equation}
  \Gamma_{0}=\gamma_{0}=q \mathbf{A}(\mathbf{X})\cdot d\mathbf{X}+p_{\parallel}\mathbf{b}(\mathbf{X})\cdot d\mathbf{X}-wdt
\end{equation}
\begin{equation}
H_{0}=h_{0}=mc^{2}\sqrt{1+p_{\parallel}^{2}/m^{2}c^{2}}-w
\end{equation}
Here all the terms are evaluated in the guiding-center coordinates. To the zeroth order, we adopt $\mathbf{X}=\mathbf{x}$ in the present study. 

The first order term in the guiding-center one-form can be obtained as
\begin{equation}
\Gamma_{1}=\gamma_{1}-(\mathcal{L}_{\mathbf{G}_{1}}\gamma_{0})_{1}+dS_{1}
\end{equation}
\begin{equation}
H_{1}=-\mathcal{L}_{\mathbf{G}_{1}} h_{0}
\end{equation}
where $S_{1}$ is the gauge function. $(\mathcal{L}_{\mathbf{G}_{1}}\gamma_{0})_{1}$ is the term in $(\mathcal{L}_{\mathbf{G}_{1}}\gamma_{0})$ with the order of $\epsilon$. Direct calculation of the the Lie derivative reveals that
\begin{eqnarray}
\label{eq:iG-gamma0}
\mathcal{L}_{\mathbf{G}}\gamma_{0}=&q \mathbf{B}\times \mathbf{G}^{X}\cdot d\mathbf{X}
+G^{p_{\parallel}}\mathbf{b}\cdot d\mathbf{X} -\mathbf{G}^{X}\cdot\mathbf{b}dp_{\parallel}-p_{\parallel}\mathbf
G^{X} \times\nabla\times \mathbf{b} \cdot d\mathbf{X}\nonumber\\
&+d(i_{\mathbf{G}}\cdot \gamma_{0}),
\end{eqnarray}
\begin{equation}
  \mathcal{L}_{\mathbf{G}} h_{0}=\frac{p_{\parallel}/m}{\sqrt{1+p_{\parallel}^{2}/m^{2}c^{2}}}G^{p_{\parallel}}.
\end{equation}
Note that the last term in Eq.~(\ref{eq:iG-gamma0}) is an exact derivative and can be combined into the gauge term $d S_{1}$.
According to the ordering $\rho_\parallel\ll L$, the fourth term $p_{\parallel}\mathbf
G^{X} \times\nabla\times \mathbf{b}$ on the right-hand-side is one order of magnitude smaller than the first term $q \mathbf{B}\times \mathbf{G}^{X}$. Therefore for the Lie derivative $\mathcal{L}_{\mathbf{G}_{1}}\gamma_{0}$, we will ignore the fourth term. The fourth term will be denoted as $(\mathcal{L}_{\mathbf{G}_{1}}\gamma_{0})_{2}$ and included $\Gamma_{2}$ instead.
%\eero{After the initial ordering, you are not allowed to do this anymore. The first term is to cancel the perpendicular term in $\gamma_1$, so you end up with nothing to compare with in the end result. Instead, you must use the gauge function to kill any $\theta$ dependency in $G^{p_{\parallel}}$. Or you set it to zero, if it does not cause further trouble.}

Note that the $\mathbf{p}_{\perp}$ term in $\gamma_{1}$ is a rapid oscillatory term. To make $\Gamma_{1}$ independent of gyro motion, we can choose
\begin{equation}
\mathbf{G}_{1}^{X}=-\frac{\mathbf{p}_{\perp}\times\mathbf{b}}{qB}
\end{equation}
to cancel the $\mathbf{p}_{\perp}$ term term. However, this is not the unique choice of $\mathbf{G}_{1}^{X}$. The term $\mathbf{G}_{1}^{X}$ can also contain other terms that are independent of the gyro motion. Here we choose $\mathbf{G}_{1}^{X}$ as
\begin{equation}
\mathbf{G}_{1}^{X}=-\frac{(\mathbf{p}_{\perp}-\mathbf{p}_{\perp a})\times\mathbf{b}}{qB},
\end{equation}
where
\begin{equation}
\label{eq:p-perpa}
\mathbf{p}_{\perp a}=-p_{\parallel}^{2}\frac{\bm{\kappa}\times\mathbf{b}}{qB},
\end{equation}
which is of the order $\epsilon$ (the same as $\mathbf{p}_{\perp}$). This transformation eliminates the gyro motion dependence in $\Gamma_{2}$, thus the other components of $\mathbf{G}_{1}$ are not necessary and can be set to zero. In terms of this, $\Gamma_{1}$ and $H_{1}$ can be written as
\begin{equation}
\Gamma_{1}=\mathbf{p}_{\perp a}\cdot d\mathbf{X}+d\sigma_{1},
\end{equation}
\begin{equation}
  H_{1}=0,
\end{equation}
where $d\sigma_{1}$ includes the derivative of the gauge function, $dS_{1}$, and the other exact derivative terms.

The second order term in the guiding-center one-form can be calculated as
\begin{eqnarray}
\Gamma_{2}&=\gamma_{2}-(L_{\mathbf{G}_{1}}\gamma_{0})_{2}-L_{\mathbf{G}_{2}}\gamma_{0}-L_{\mathbf{G}_{1}}\gamma_{1}+\frac{1}{2}L^{2}_{\mathbf{G}_{1}}\gamma_{0}+dS_{2},\nonumber\\
&=\gamma_{2}-(L_{\mathbf{G}_{1}}\gamma_{0})_{2}-L_{\mathbf{G}_{2}}\gamma_{0}
-\frac{1}{2}L_{\mathbf{G}_{1}}\gamma_{1}-\frac{1}{2}L_{\mathbf{G}_{1}}\Gamma_{1}+dS'_{2},\\
  H_{2}&=h_{2}-L_{\mathbf{G}_{2}} h_{0},
\end{eqnarray}
where $S'_{2}=S_{2}-(1/2)i_{\mathbf{G}_{1}}d S_{1}$.
Note that in this expression, the $L_{\mathbf{G}_{1}}\Gamma_{1}$ term will be of the order $\epsilon^{3}$, since both $\mathbf{G}_{1}^{X}$ and $\Gamma_{1}$ are order $\epsilon$, and the Lie derivative will introduce another spatial gradient (Here we choose $\mathbf{G}_{1}$ to only contain $\mathbf{G}_{1}^{X}$).
We first consider the second term $(L_{\mathbf{G}_{1}}\gamma_{0})_{2}$, which is the fourth term in Eq.~(\ref{eq:iG-gamma0}) that we dropped previously,
\begin{equation}
\label{eq:G1X-nabla-b}
  -p_{\parallel}\mathbf{G}_{1}^{X}\times\nabla\times \mathbf{b}\cdot d\mathbf{X}=-p_{\parallel}\tau\mathbf{G}_{1}^{X}\times \mathbf{b}\cdot d\mathbf{X}-p_{\parallel}\mathbf{G}_{1}^{X}\cdot\bm{\kappa} \mathbf{b}\cdot d\mathbf{X},
\end{equation}
where $\tau=\mathbf{b}\cdot\nabla\times \mathbf{b}$.
The third term $L_{\mathbf{G}_{2}}\gamma_{0}$ can be calculated similar to Eq.~(\ref{eq:iG-gamma0}), and only the leading order terms are kept. For the fourth term,
\begin{eqnarray}
\mathcal{L}_{\mathbf{G}_{1}}d\gamma_{1}=&-\mathbf{G}_{1}^{X}\times\nabla\times \mathbf{p}_{\perp}\cdot d\mathbf{X}-\frac{\partial \mathbf{p}_{\perp}}{\partial p_{\parallel}}\cdot \mathbf{G}_{1}^{X} d p_{\parallel}-\frac{\partial \mathbf{p}_{\perp}}{\partial \mu}\cdot \mathbf{G}_{1}^{X} d\mu\nonumber\\
&-\frac{\partial \mathbf{p}_{\perp}}{\partial \theta}\cdot \mathbf{G}_{1}^{X}d\theta+d(i_{\mathbf{G}_{1}}\cdot d\gamma_{1}).
\end{eqnarray}
Note that the first term will be of the order $\epsilon^{3}$, since both $\mathbf{G}_{1}^{X}$ and $\mathbf{p}_{\perp}$ are of the order $\epsilon$, and the spatial gradient will introduce another $\epsilon$. Therefore for $\Gamma_{2}$ we will consider the other terms.

Following the above derivations, the second-order guiding-center one-form can be written as
\begin{eqnarray}
\Gamma_{2}&=&-(q\mathbf{B}\times \mathbf{G}_{2}^{X}+p_{\parallel}\tau \mathbf{G}_{1}^{X}\times\mathbf{b})\cdot d\mathbf{X}-(G_{2}^{p_{\parallel}}+p_{\parallel}\mathbf{G}_{1}^{X}\cdot\bm{\kappa}) \mathbf{b}\cdot d\mathbf{X}\nonumber\\
&&-\left(\mathbf{G}_2^{X}\cdot \mathbf{b}-\frac{1}{2}\frac{\partial \mathbf{p}_{\perp}}{\partial p_{\parallel}}\cdot \mathbf{G}_{1}^{X} \right)d p_{\parallel}+\frac{1}{2}\frac{\partial \mathbf{p}_{\perp}}{\partial \mu}\cdot \mathbf{G}_{1}^{X} d\mu\nonumber\\
&&+\frac{1}{2}\frac{\partial \mathbf{p}_{\perp}}{\partial \theta}\cdot \mathbf{G}_{1}^{X}d\theta\label{eq:Gamma2},\\
H_{2}&=&\frac{p_{\parallel}/m}{\sqrt{1+p_{\parallel}^{2}/m^{2}c^{2}}}G_{2}^{p_{\parallel}}+\frac{p_{\perp}^{2}}{2m\sqrt{1+p_{\parallel}^{2}/m^{2}c^{2}}}\label{eq:H2}.
\end{eqnarray}
To ensure that the term with $\mathbf{b}\cdot d\mathbf{X}$ in $\Gamma_{2}$ does not depend on gyro-motion, we can choose
\begin{equation}
  G_{2}^{p_{\parallel}}=-p_{\parallel}\mathbf{G}_{1}^{X}\cdot \bm{\kappa}.
\end{equation}
This choice will affect the Hamiltonian $H_{2}$, which involves two terms that both depend on the gyro motion. However, using Eq.~(\ref{eq:mu}) and Eq.~(\ref{eq:p-perpa}), $H_{2}$ can be written as,
\begin{equation}
H_{2}=\frac{\mu B}{\sqrt{1+p_{\parallel}^{2}/m^{2}c^{2}}}+\frac{p_{\perp a}^{2}}{2m\sqrt{1+p_{\parallel}^{2}/m^{2}c^{2}}},
\end{equation}
which becomes independent on $\theta$.

%However, we already know that $G_{2}^{p_{\parallel}}$ is an oscillatory term, and there is no other terms from Lie transformation to condensate that. In terms of that, the conservation of $H_{2}$ leads to the fact that $p_{\perp}^{2}/2m$ is not a conserved quantity in the gyro motion but a fast oscillating one. This is actually consistent with the numerical simulation result. The conserved quantity instead is
%\begin{equation}
%  p_{\perp}^{2}+2p_{\parallel}^{2}\frac{\kappa\times \mathbf{b}}{qB}\cdot\mathbf{p}_{\perp}
%\end{equation}
%or
%\begin{equation}
%\label{eq:conservation}
% \left\lvert \mathbf{p}_{\perp}+p_{\parallel}^{2}\frac{\kappa\times %\mathbf{b}}{qB}\right\rvert^{2}
%\end{equation}
%where we didn't put the $(\mathbf{G}_{1}^{X})_{0}$ term since it is independent of gyro motion.

We now consider the $d\mu$ and $d\theta$ terms in $\Gamma_{2}$.
According to Eq.~(\ref{eq:mu}) and Eq.~(\ref{eq:p-perpa}), $\mu$ measures the absolute value  $|\mathbf{p}_{\perp}-\mathbf{p}_{\perp a}|$. In terms of that, we can choose $\theta$ as the slope angle for the vector $\mathbf{p}_{\perp}-\mathbf{p}_{\perp a}$. The value of $\mathbf{p}_{\perp}$ can be expressed as
\begin{equation}
  \mathbf{p}_{\perp}=\mathbf{p}_{\perp a}+\sqrt{2m\mu B}\hat{\perp}(\mathbf{X}, \theta),
\end{equation}
where $\hat{\perp}(\mathbf{X}, \theta)$ is the unit vector characterizing the direction of $\mathbf{p}_{\perp}-\mathbf{p}_{\perp a}$. This relationship yields
\begin{equation}
  \frac{\partial\mathbf{p}_{\perp}}{\partial \theta}\cdot \mathbf{G}_{1}^{X}=(\mathbf{p}_{\perp}-\mathbf{p}_{\perp a})\times \mathbf{b}\cdot \mathbf{G}_{1}^{X}=\frac{2m\mu}{q},
\end{equation}
and
\begin{equation}
  \frac{\partial\mathbf{p}_{\perp}}{\partial \mu}\cdot \mathbf{G}_{1}^{X}=\frac{\mathbf{p}_{\perp}-\mathbf{p}_{\perp a}}{2\mu}\cdot \mathbf{G}_{1}^{X}=0.
\end{equation}
For the $d\mathbf{X}$ and $d p_{\parallel}$ terms in Eq.~(\ref{eq:Gamma2}), we can choose
\begin{equation}
  \mathbf{G}_{2}^{X}=\frac{\tau p_{\parallel}}{qB}\mathbf{G}_{1}^{X}-\frac{p_{\parallel}\bm{\kappa}\times\mathbf{b}}{qB}\cdot \mathbf{G}_{1}^{X} \mathbf{b},
\end{equation}
to make both terms vanish.
After these simplifications, $\Gamma_{2}$ can be rewritten as
\begin{equation}
  \Gamma_{2}=\frac{m\mu}{q}d\theta.
\end{equation}
Thus
\begin{eqnarray}
  \Gamma&=\Gamma_{0}+\epsilon \Gamma_{1}+\epsilon^{2} \Gamma_{2}+\dots\nonumber\\
  &=q \mathbf{A}\cdot d\mathbf{X}+p_{\parallel}\mathbf{b}\cdot d\mathbf{X}-\epsilon p_{\parallel}^{2}\frac{\bm{\kappa}\times \mathbf{b}}{qB}\cdot d\mathbf{X}+\epsilon^{2}\frac{m\mu}{q}d\theta-wdt+\dots\nonumber\\
  H&=H_{0}+\epsilon H_{1}+\epsilon^{2} H_{2}+\dots\nonumber\\
  &=\gamma_{r} mc^{2}-w+\dots
\end{eqnarray}
where
\begin{equation}
\gamma_{r}=\sqrt{1+\frac{p_{\parallel}^{2}\left[1+\epsilon^{2} p_{\parallel}^{2}\kappa^{2}/(q^{2}B^{2})\right]+2\epsilon^{2} m\mu B}{m^{2}c^{2}}}
\end{equation}
is the Lorentz factor in the guiding-center coordinates. In this derivation, we put the terms into the square root by introducing correction terms of higher order.

In the above derivation we didn't include the effects  of electric potential in the phase-space-time one form. Although the acceleration from electric force is very important for REs, the acceleration is mainly due to the inductive electric field instead of static one, which can be described by the change of the magnetic field vector potential with time, $\partial \mathbf{A}/\partial t$. The static electric field and the potential energy associated with it can be regarded as a high order term in the particle Hamiltonian, compared to the kinetic energy.  Therefore we can set the electric potential energy, $q\Phi$, as a high-order term $\epsilon^{2}$ in the guiding-center Hamiltonian,
\begin{eqnarray}
  H=\gamma_{r} mc^{2}+q\Phi-w+\dots.
\end{eqnarray}

\section{Equation of motion of guiding-center coordinates}
\label{sec:equation-of-motion}

Using the symplectic part of the phase-space-time one form, we can derive the guiding-center Poisson bracket for relativistic runaway electrons~\cite{cary_hamiltonian_2009,hirvijoki_guiding-centre_2015},
\begin{eqnarray}
  \{F,G\}_{gc}=&\frac{q}{m}\left(\frac{\partial F}{\partial \theta}\frac{\partial G}{\partial \mu}-\frac{\partial F}{\partial \mu}\frac{\partial G}{\partial \theta}\right)\nonumber\\
  &+\frac{\mathbf{B}^{*}}{B_{\parallel}^{*}}\cdot \left(\nabla^{*}\frac{\partial G}{\partial p_{\parallel}}-\frac{\partial F}{\partial p_{\parallel}}\nabla^{*} G\right)-\frac{\mathbf{b}^{*}}{q B^{*}_{\parallel}}\cdot \nabla^{*} F\times\nabla^{*} G\nonumber\\
  &+\left(\frac{\partial F}{\partial w}\frac{\partial G}{\partial t}-\frac{\partial F}{\partial t}\frac{\partial G}{\partial w}\right),
\end{eqnarray}
where
\begin{equation}
\mathbf{A}^{*}=\mathbf{A}+\frac{p_{\parallel}}{q}\mathbf{b}-p_{\parallel}^{2}\frac{\bm{\kappa}\times \mathbf{b}}{q^{2}B},
\end{equation}
\begin{equation}
  \mathbf{B}^{*}=\nabla\times\mathbf{A}^{*},
\end{equation}
\begin{equation}
  \mathbf{b}^{*}=\mathbf{b}-2p_{\parallel}\frac{\bm{\kappa}\times \mathbf{b}}{qB},
\end{equation}
\begin{equation}
  B^{*}_{\parallel}=\mathbf{B}^{*}\cdot \mathbf{b}^{*}.
\end{equation}
\begin{equation}
  \nabla^{*}=\nabla-q\frac{\partial \mathbf{A^{*}}}{\partial t}\frac{\partial}{\partial w}
\end{equation}

Base on that, we can obtain the Hamiltonian equation of motion for each coordinate. We find that, to the leading order,
\begin{equation}
  \dot{\mathbf{X}}=\{\mathbf{X},H\}_{gc}=\frac{p_{\parallel}^{\star}}{\gamma_{r} m}\frac{\mathbf{B}^{*}}{B^{*}_{\parallel}}+\frac{\mathbf{b}^{*}}{qB^{*}_{\parallel}}\times\left(\nabla H+\frac{\partial \mathbf{A}^{*}}{\partial t}\right),
\end{equation}
\begin{equation}
  \dot{p_\parallel}=\{p_{\parallel},H\}_{gc}=-\frac{\mathbf{B}^{*}}{B^{*}_{\parallel}}\cdot\left(\nabla H+\frac{\partial \mathbf{A}^{*}}{\partial t}\right),
\end{equation}
\begin{equation}
  \dot{\mu}=\{\mu,H\}_{gc}=0,
\end{equation}
\begin{equation}
  \dot{\theta}=\{\theta,H\}_{gc}=\frac{qB}{\gamma_{r} m},
\end{equation}
where
\begin{equation}
  p_{\parallel}^{\star}=p_{\parallel}+2p_{\parallel}^{3}\frac{\bm{\kappa}^{2}}{q^{2}B^{2}},
\end{equation}
\begin{equation}
  \nabla H=\frac{1}{\gamma_{r}} \left(\mu\nabla B+ \frac{p_{\parallel}^{4}}{2mq^{2}}\nabla\frac{\bm{\kappa}^2}{B^{2}}\right)+q\nabla \Phi.
\end{equation}

Note that both the Possion bracket and the equations of motion are very similar to those of standard guiding-center coordinates~\cite{cary_hamiltonian_2009}, and the difference are all higher order corrections. This means that most of the established frameworks of guiding-center simulation can be applied to runaway electron studies in the lower energy regime (so that the guiding-center ordering still holds) with slight modifications, and can be used to study the runaway electron dynamics in tokamaks like the case in Sec. \ref{sec:numerical-indication}. However, for highly energetic runaway electrons, the guiding-center ordering $\rho_{\parallel}\ll L$ will not hold any more because of the Lorentz factor. In this case, one has to rely on full-orbit simulation model to study the dynamics of runaway electrons.
 
\section{Comparison with higher order terms of magnetic moment in standard guiding-center theory}
\label{sec:higher-order-terms}

It is realized in the derivation in Sec. \ref{sec:guiding-center-transformation} that the difference of the new guiding-center coordinates from the standard ones, including the new magnetic moment, is caused by a new ordering $p_{\perp}\sim \epsilon p_{\parallel}$. In the standard derivation of guiding-center coordinates, the $\mu_{1}$ and $\mu_{2}$ terms in Eq.~(\ref{eq:mu}) will only appear in the higher order corrections to $\mu$.
In literature, there are very few instances of an explicit expression for the higher order correction to the magnetic moment. For a nonrelativistic particle such an expression can be found in Ref. \cite{burby_automation_2013}. It is likely that the expression can be generalized to be applicable for a relativistic particle by simply replacing any $m\mathbf{v}$ with $\mathbf{p}$. 

As shown in \cite{burby_automation_2013}, the second order correction to $\mu$ for arbitrary $\mathbf{p}$ and a general magnetic field is very complicated. However, after applying the runaway electron ordering ($p_{\perp}\sim \epsilon p_{\parallel}$), we find that most of the terms in $\mu_{1}$ and $\mu_{2}$ expressions in \cite{burby_automation_2013} are actually of $O(\epsilon^{3})$ or higher order, and the expression can be greatly simplified by omitting them.

By introducing the new ordering, the second order term in the $\mu_1$ expression in \cite{burby_automation_2013} is
\begin{eqnarray}
\mu_{1}&=\frac{1}{qm|B|^{2}}\left(\frac{1}{4}\mathbf{p}\cdot\nabla \mathbf{b}\cdot (\mathbf{p}\times \mathbf{b})(\mathbf{p}\cdot \mathbf{b})-\frac{5}{4}\mathbf{b}\times\bm{\kappa}\cdot \mathbf{p} (\mathbf{p}\cdot \mathbf{b})^{2}\right),\nonumber\\
&=\frac{\mathbf{p}_{\perp}\cdot (\bm{\kappa}\times\mathbf{b}) p_{\parallel}^{2}}{qmB^{2}}+O(\epsilon^{3}).
\end{eqnarray}

The second order term in the $\mu_2$ in \cite{burby_automation_2013} is
\begin{eqnarray}
\mu_2&=\frac{1}{mq^{2}}\left[\frac{p_{\parallel}}{6|B|^{3}}\mathbf{p}\mathbf{p}:\nabla\nabla (\mathbf{b}\cdot \mathbf{p})-\frac{29p_{\parallel}^{2}}{24|B|^{3}}\bm{\lambda}\cdot \bm{\lambda}+\frac{5 p_{\parallel}^{2}}{12|B|^{3}}\mathbf{b}\mathbf{p}:\nabla\nabla(\mathbf{b}\cdot \mathbf{p})\right.\nonumber\\
&\left.+\frac{5p_{\parallel}^{3}}{3|B|^{3}}\bm{\lambda}\cdot\bm{\kappa}+\frac{5p_{\parallel}^{3}}{12|B|^{3}}\mathbf{b}\mathbf{b}:\nabla\nabla(\mathbf{b}\cdot \mathbf{p})+\frac{25p_{\parallel}^{4}}{24|B|^{3}}\bm{\kappa}\cdot\bm{\kappa}\right], \nonumber\\
&=\frac{p_{\parallel}^{4}\kappa^2}{2mq^{2}B^{3}}+O(\epsilon^{3}).
\end{eqnarray}
where $\bm{\lambda}=\mathbf{p}\cdot\nabla\mathbf{b}$, and we have used the equation
\begin{equation}
\mathbf{b}\cdot (\mathbf{b}\mathbf{b}:\nabla\nabla) \mathbf{b}=-\bm{\kappa}\cdot\bm{\kappa}.
\end{equation}
This equation can be proved using the fact that $\mathbf{b}\cdot\mathbf{b}=1$ is a constant. Therefore, with a small parameter $\alpha$,
\begin{eqnarray}
\mathbf{b}(\mathbf{X}+\alpha \mathbf{b})^{2}=&\mathbf{b}(\mathbf{X})^{2}+2\alpha \mathbf{b}\cdot (\mathbf{b}\cdot\nabla)\mathbf{b}+\alpha^2 \mathbf{b}\cdot (\mathbf{b}\mathbf{b}:\nabla\nabla) \mathbf{b}\nonumber\\
&+\alpha^2 [(\mathbf{b}\cdot\nabla)\mathbf{b}]^{2}+O(\alpha^3).
\end{eqnarray}
Examining the terms with $\alpha$ and $\alpha^2$ reveals that
\begin{equation}
\mathbf{b}\cdot \bm{\kappa}=0,
\end{equation}
\begin{equation}
\mathbf{b}\cdot (\mathbf{b}\mathbf{b}:\nabla\nabla) \mathbf{b}+\bm{\kappa}\cdot\bm{\kappa}=0.
\end{equation}

Combining $\mu_0$ and $\mu_{1}$, $\mu_{2}$,
\begin{equation}
\mu_0+\mu_{1}+\mu_{2}=\frac{1}{2mB}\left\vert \mathbf{p}_{\perp}+p_{\parallel}^{2}\frac{\bm{\kappa}\times \mathbf{b}}{qB}\right\vert^{2}.
\end{equation}
which is the same as Eq.~(\ref{eq:mu}).

In this section we only consider the terms of $\epsilon^{2}$ in $\mu$ in the runaway electron guiding-center ordering. For highly relativistic runaway electrons, the value of $\epsilon$ becomes larger, and the higher order terms can become more important. This is the reason that the value of $\mu$ can deviate from its original value in the later time in Fig.~\ref{fig:mu-conservation3s}. The significance of these high order correction terms indicates the breakdown of the guiding-center approximation, as discussed in \cite{liu_collisionless_2016}.

\section{Conclusion}
\label{sec:conclusion}

In this paper we show that for runaway electrons, the breakdown of lowest magnetic magnetic moment in standard guiding-center theory as an adiabatic invariant can be partly addressed by introducing a new expression of magnetic moment, assuming that the guiding-center ordering is still valid. The new magnetic moment includes correction terms depending on $p_{\parallel}$ and the magnetic field curvature, which can be found in the higher order corrections of $\mu$ in the standard guiding-center theory. Using the fact that runaway electrons have anisotropic distribution in momentum space ($p_{\perp}\ll p_{\parallel}$), we successfully derived a new set of guiding-center coordinates and the corresponding phase-space-time one-form for runaway electrons, including the new expression for magnetic moment. In addition, with the help of a full-orbit particle simulation model, we show that the new expression for magnetic moment is conserved much better than the standard magnetic moment.

Using the new expression for $\mu$ and assuming it is a good invariant, we explain the collisionless pitch-angle scattering found previously~\cite{liu_collisionless_2016,carbajal_space_2017}. This explanation does not violate the guiding-center ordering, implying that a simulation model based on the guiding-center approximation can still be applied to study runaway electrons in certain cases. Using the non-canonical Hamiltonian approach, we derived a new Poisson bracket and a new set of equations of motion for guiding-center coordinates, which can be used to develop a guiding-center simulation framework for runaway electrons. However, the work presented will be invalid for runaway electron with extremely high energy ($>80$MeV), for which the presented second-order guiding-center theory breaks down. In this case a full-orbit particle simulation is required.

In addition to the breakdown of guiding-center approximation, the conservation of magnetic moment can also be violated through dissipative forces, including the collisions and radiation forces. For relativistic runaway electrons, the collisional effects are weak, but the radiation effects including synchrotron radiation~\cite{martin-solis_momentumspace_1998,stahl_effective_2015,aleynikov_theory_2015,hirvijoki_radiation_2015} and bremsstrahlung~\cite{bakhtiari_role_2005,embreus_effect_2016}  can be significant since the radiation power increases with particle energy. Fortunately, the radiation effects can be addressed within the guiding-center framework, by transforming the radiation forces from particle coordinates to guiding-center coordinates~\cite{hirvijoki_guiding-centre_2015}. Note that in~\cite{hirvijoki_guiding-centre_2015}, the radiation reaction effects due to particle gyro-motion is transformed to guiding-center coordinates, but the radiation effects due to magnetic field curvature is missing since the transformation is only taken to the first order. In order to capture this effect, one can use the runaway electron guiding-center ordering in Eq.~(\ref{eq:ordering}) and the guiding-center coordinates introduced in this paper to transform the radiation force. This will be discussed in future.

\ack

Chang Liu wants to thank Joshua W. Burby and Alain J. Brizard for fruitful discussions.
This work has received funding from
the Department of Energy 
under Grant No. DE-SC0016268 and DE-AC02-09CH11466.

\appendix

\section{Proof of uniqueness of $\mu$}

In the previous derivations, we first introduce the definition of $\mu$ (Eq.~(\ref{eq:mu})) and then show it is a good candidate of magnetic moment. In this section, we illustrates that in order to get the desired form of $\Gamma$, this choice of $\mu$ is unique.

We start from the $d\mu$ and $d\theta$ terms in Eq.~(\ref{eq:Gamma2}). To have the desired expression in the one-form, we want
\begin{equation}
    \frac{\partial\mathbf{p}_{\perp}}{\partial \mu}\cdot \mathbf{G}_{1}^{X}=0,\qquad \frac{\partial\mathbf{p}_{\perp}}{\partial \theta}\cdot \mathbf{G}_{1}^{X}=\frac{2m\mu}{q}.
\end{equation}

We first decide the definite of $\theta$. We choose $\theta$ as the slope angle of the vector $\mathbf{p}_{\perp}-\mathbf{p}_{\perp b}$, where $\mathbf{p}_{b}$ is a variable depending on $p_{\parallel}$ and $\mathbf{X}$. In this case,
\begin{equation}
\frac{\partial\mathbf{p}_{\perp}}{\partial \mu}\cdot \mathbf{G}_{1}^{X}=\frac{\partial |\mathbf{p}_{\perp}-\mathbf{p}_{\perp b}|}{\partial \mu}\frac{\mathbf{p}_{\perp}-\mathbf{p}_{\perp b}}{|\mathbf{p}_{\perp}-\mathbf{p}_{\perp b}|}\cdot \frac{(\mathbf{p}_{\perp}-\mathbf{p}_{\perp a})\times\mathbf{b}}{qB}=0.
\end{equation}
To satisfy this condition, it is required that $\mathbf{p}_{\perp b}=\mathbf{p}_{\perp a}$.

Given that, the expression of $\mu$ is
\begin{equation}
\label{eq:mu2}
  \mu=\frac{q}{2m}\frac{\partial\mathbf{p}_{\perp}}{\partial \theta}\cdot \mathbf{G}_{1}^{X}=\frac{|\mathbf{p}_{\perp}-\mathbf{p}_{\perp a}|^{2}}{2mB}.
\end{equation}

Then we turn to the expression of $H_{2}$ in Eq.~(\ref{eq:H2}).
\begin{equation}
\label{eq:H2x}
H_{2}=\frac{1}{\sqrt{1+p_{\parallel}^{2}/m^{2}c^{2}}}\left[\frac{p_{\parallel}^{2}}{qBm}(\bm{\kappa}\times \mathbf{b})\cdot(\mathbf{p}_{\perp}-\mathbf{p}_{\perp a})+\frac{1}{2m}p_{\perp}^{2}\right].
\end{equation}
We know that in order to have the desired form of guiding-center Hamiltonian, $H_{2}$ should be expressed as $a\mu B+b$, where $a$ and $b$ are $\theta$ independent. In terms of that, we compare the $\mathbf{p}_{\perp}$ term in Eq.~(\ref{eq:mu2}) and Eq.~(\ref{eq:H2x}). We found that the only choice that can satisfy this condition is
\begin{equation}
  \mathbf{p}_{\perp a}=-p_{\parallel}^{2}\frac{\bm{\kappa}\times\mathbf{b}}{qB},
\end{equation}
which is the choice we made in the main text. Then we can obtain Eq.~(\ref{eq:mu}) according to Eq.~(\ref{eq:mu2}).

\section{Derivation of guiding-center one-form using initial Lie transform}

An alternative approach to derive the guiding-center one-form for relativistic runaway electrons is to apply a zeroth order Lie transform at the particle one-form initially, before the guiding-center approximation. The initial Lie transform can be expressed as
\begin{equation}
  T^{-1}_{0}=\exp \left(-\mathcal{L}_{\mathbf{G}_{0}}\right)
\end{equation}
where
\begin{equation}
\label{eq:G0p}
\bm{G}_0^{\bm{x}}=0 \qquad \bm{G}^{\bm{p}}_0=-\frac{(\bm{p}\cdot\bm{b})^2}{qB}\bm{\kappa}\times\bm{b}.
\end{equation}
The transformed particle one-form is
\begin{equation}
  T^{-1}_{0}\gamma=q\bm{A}\cdot d\bm{x}+(p_{\parallel}\bm{b}+\bm{p}_{\bot})\cdot d\bm{x}-\frac{p_{\parallel}^2}{qB}\bm{\kappa}\times\bm{b}\cdot d\mathbf{x}-wdt
\end{equation}
and the Hamiltonian is
\begin{equation}
\label{eq:T0mu}
  T^{-1}_{0}h=mc^{2}\gamma-\frac{p_{\parallel}^2}{\gamma mqB}\bm{\kappa}\times\bm{b}\cdot\bm{p}_{\bot}+\frac{1}{2}\frac{p_{\parallel}^4}{\gamma mq^2B^2}\left(|\bm{\kappa}\times\bm{b}|^2-\frac{(\bm{\kappa}\times\bm{b}\cdot\bm{p}_{\bot})^2}{\gamma^2m^2c^2}\right)-w.
\end{equation}

Based on the new $\gamma$ and $h$, one can use Lie-perturbation method to find the guiding-center one-form and Hamiltonian, following the ordering in Eq.~(\ref{eq:ordering}) and the steps in Sec. \ref{sec:guiding-center-transformation}. Then in the derivation of the second-order terms in the guiding-center Hamiltonian, it is found that the correction term brought by $G_{2}^{p_{\parallel}}$, as shown in Eq.~(\ref{eq:H2}), can be canceled by the second term on the right-hand-side of Eq.~(\ref{eq:T0mu}), which makes the Hamiltonian independent of the gyro-motion. In other words, with the help of the initial transformation, there is no need to introduce a new definition of magnetic moment, and the the guiding-center transformation becomes the same as the standard one.

The magnetic moment is simply $\mu=p_{\perp}^{2}/2mB$ in the transformed particle coordinates. Considering Eq.~(\ref{eq:G0p}), the expression in the original particle coordinates is
\begin{equation}
\mu_p=\frac{1}{2mB}\left|\bm{p}_{\bot}+\frac{(\bm{p}\cdot\bm{b})^2}{qB}\bm{\kappa}\times\bm{b}\right|^2,
\end{equation}
which is the same as Eq.~(\ref{eq:mu}). This approach is similar to the non-perturbation transform used in \cite{burby_toroidal_2017} to simplify the Lagrangian, but here we apply the transform on the particle Lagrangian and Hamiltonian before guiding-center transform instead of after it.

%#\input{eero_notes}

\bibliographystyle{iopart-num}
\bibliography{magneticmoment}

\end{document}